\documentclass[aps, prl, amsmath,amssymb, reprint, showpacs,superscriptaddress,groupedaddress, nofootinbib ]{revtex4-1}

\usepackage{graphicx}
\usepackage{dcolumn}
\usepackage{bm}
\usepackage{amsmath}
\usepackage{color}

\usepackage{dcolumn}  
\usepackage{amsmath}
\usepackage{multirow}
\usepackage{graphicx}   
\usepackage{subfigure}
\usepackage{comment}
\usepackage{color}
\usepackage{nicefrac}
\usepackage{soul}
\usepackage{slashbox}
\usepackage{bm,array}
\usepackage{tikz}
\usepackage{pifont}
\usepackage{gensymb}
\usepackage{float} 
\newcolumntype{C}{>{\centering\arraybackslash}p{1em}}

\begin{document}
	\title{Optoelectronic and Transport Properties of Vacancy Ordered Double Perovskite Halides: A First-principles Study  } 
	\author{ Supriti Ghorui$^1$, Jiban Kangsabanik$^{1,2}$, M. Aslam$^1$ and  Aftab Alam$^{1,*}$}
	\affiliation{$^1$Department of Physics, Indian Institute of Technology Bombay, Powai, Mumbai 400076, India}
 \affiliation{$^2$Department of Physics, Technical University of Denmark, 2800 Kongens Lyngby, Denmark}
\email{aftab@iitb.ac.in}
	
	\begin{abstract}
In the search for stable lead (Pb) free perovskites, Vacancy ordered double perovskite (VODP), A$_2$BX$_6$ has emerged as a promising class of  materials for solar harvesting owing to their nontoxicity, better stability, and unique optoelectronic properties. Recently, this class has been explored for a wide range of applications such as photovoltaics, photodetectors, photocatalysis, and light-emitting diodes. Here, we present the stability and the key physical attributes of few selected compounds in a systematic manner using state-of-the-art first-principle calculations. A careful structural and stability analysis via simulating convex hull and compositional phase diagrams for different structural prototypes discloses 14 stable and 1 metastable compounds in this class. The electronic structure calculations using hybrid functional reveals six compounds to acquire band gap in the ideal visible region. These six compounds, namely Cs$_2$SnI$_6$, Cs$_2$PdI$_6$, Cs$_2$TeI$_6$, Cs$_2$TiI$_6$, Cs$_2$PtI$_6$, and Cs$_2$PdBr$_6$, show high optical absorption ($\approx$ 10$^{5}$ cm $^{-1}$) giving rise to high spectroscopic limited maximum efficiency, SLME (15-23\%) in the thin-film thickness range. Close inspection of transport properties reveals polar optical phonon scattering to be the dominant mechanism limiting the overall mobility. Further analysis of the polaron excitations discloses the possibility of large polaron formation at low to moderate defect concentrations. At high defect concentrations, ionized impurity scattering takes over. This suggests that, a simulation based guided control of defect concentrations during synthesis can yield a desired candidate for promissing device application. Additionally, few selected compounds show moderate to high electron mobility values ($\sim$13-63 cm$^2$V$^{-1}$ s$^{-1}$) at room temperature. Overall, the present study paves an important path to help design VODP as Pb-free  potential candidates for future optoelectronic applications. 
\end{abstract}

	\maketitle

     \section*{I. Introduction} 
     
\begin{figure*}[t]
     	 			\centering
     	 			\includegraphics[width=.9\linewidth]{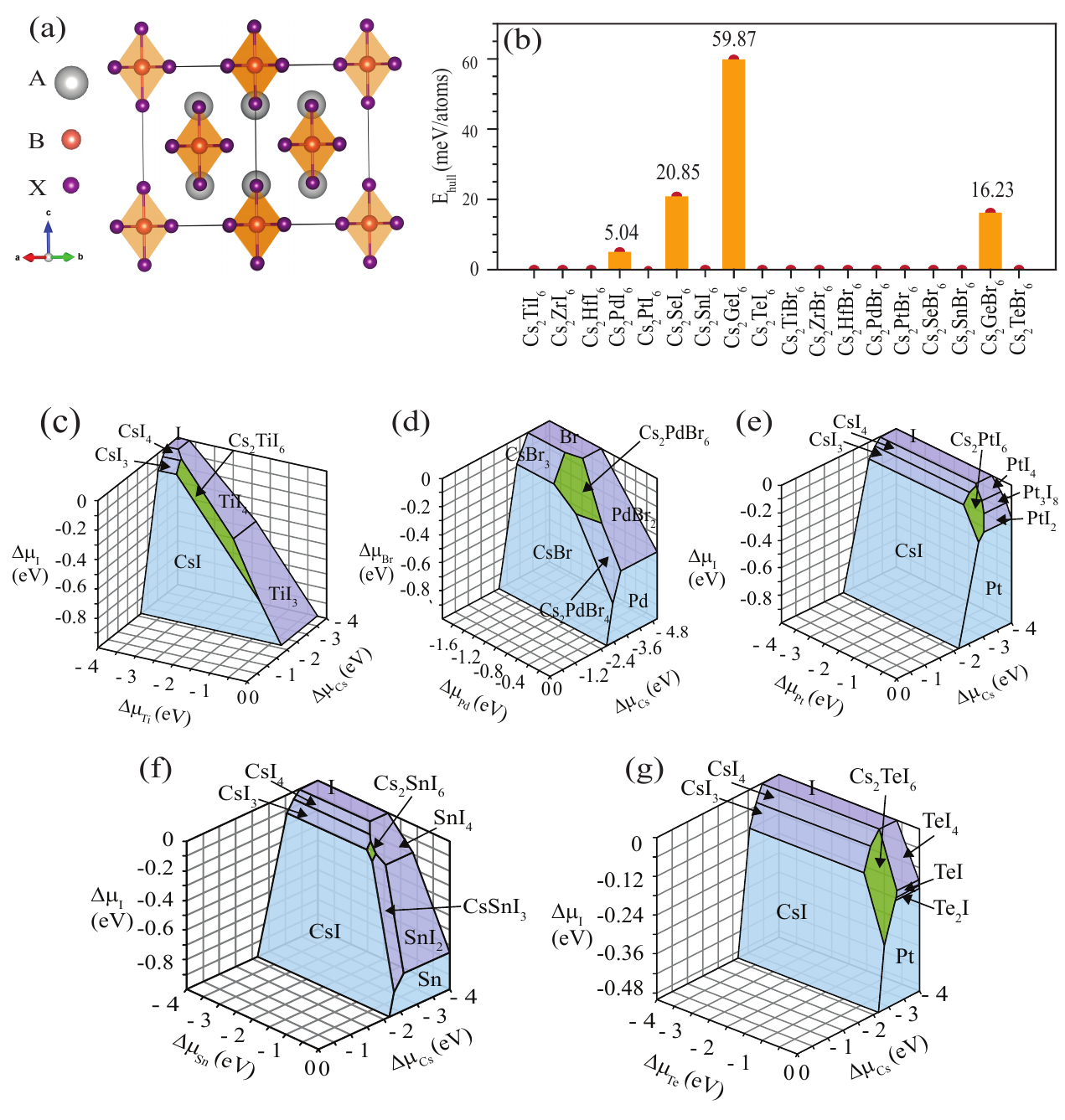}
    	 			\caption{(a) Crystal structure of  A$_2$BX$_6$ compounds in Fm-3m space group. (b)  Convex energy hull (E$_{\mathrm{hull}}$)  for Cs$_2$BX$_6$ (B= Pd, Pt, Ti, Hf, Zr, Ge, Te, Se, Sn; X=I, Br). Red circle denotes the top of the bar.  Compositional chemical potential phase diagram of (c) Cs$_2$TiI$_6$, (d) Cs$_2$PdBr$_6$, (e) Cs$_2$PtI$_6$, (f) Cs$_2$SnI$_6$, and (g) Cs$_2$TeI$_6$  with respect to competitive secondary phases. The green shaded regions show the stable regions of the corresponding materials. }
     	 			\label{fig:1}
     	 		\end{figure*}

     Lead halide perovskites(LHP) have reignited immense research interest in the Photovoltaics (PV) community due to their remarkable  power conversion efficiency(PCE) of 25.6\% \cite{jeong2021pseudo} (till date) and affordable device processability. The rapid rise in PCE (3.8\% to 25.6\%) in a short period of time (2009-2021) is  attributed to its high absorption coefficient, high charge carrier mobility, defect tolerance, and cost-effective flexible synthesis. Because of their suitable optoelectronic properties, they have also been explored as photodetectors (PD)\cite{li2019enhanced,huang2021precursor}, photocatalysts(PC) \cite{zhang2019stable,yin2021controlling}, and light emitting diodes (LED)  \cite{xu2019rational,cui2021efficient}. Yet, there remains two major challenges in their large scale scalability: (1) Lead (Pb) toxicity  and (2) stability in the ambient environment. At present, major research efforts at laboratory scale have been devoted to overcome these issues without losing their original PV performance. \cite{lin2020piperidinium, stolterfoht2018visualization,giustino2016toward,shao2018highly,kangsabanik2020optoelectronic} This has led to a detailed exploration of the diverse chemical space of halides perovskites (ABX$_3$)\cite{stoumpos2013semiconducting} and their derivatives.\cite{zhao2017design,xia2020unveiling,cai2017computational}   Among these perovskite derivatives, three major stoichiometric classes have garnered immense research interest. One of the classes namely double perovskites (DP) with stoichiometry A$_2$BB$'$X$_6$ is mainly generated via transmutation of a combination of trivalent and monovalent elements at B-sites.\cite{volonakis2016lead} For example,  Cs$_2$BiAgBr$_6$,\cite{volonakis2016lead,slavney2016bismuth,kangsabanik2018double} Cs$_2$InAgCl$_6$,\cite{volonakis2017cs2inagcl6} etc. belong to DP class which have been extensively explored for various optoelectronic applications. Similarly,  A$_3$B$_2$X$_9$ (e.g. Cs$_3$Bi$_2$I$_9$\cite{bai2018lead}, Cs$_3$Sb$_2$I$_9$\cite{singh2018photovoltaic} etc.) and A$_2$BX$_6$ (e.g. Cs$_2$SnI$_6$\cite{qiu2017unstable}, Cs$_2$TiI$_6$\cite{ju2018earth} etc.) structures are constructed by replacing with trivalent and tetravalent atoms respectively and leaving a vacant B-site. Here, A$_2$BX$_6$ is also called  vacancy ordered double perovskite where corner shared alternate BX$_6$ octahedras are removed along all three directions from the unit cell as shown in Figure \ref{fig:1}(a).

     In the past few years, vacancy-ordered double perovskite family (A$_2$BX$_6$) has gradually drawn ample attention in a wide range of optoelectronic applications owing to their better environmental durability, tunable optical and electronic properties. For example, Cs$_2$SnI$_6$  has been studied as a potential candidate in  PV\cite{lee2017solution}, LED, PD\cite{shao2020high,huang2021precursor}, and PC applications due to its direct band gap nature in the visible range( 1.1-1.62 eV), a high absorption coefficient ($\approx$10$^5$ cm$^{-1}$)\cite{qiu2017unstable},  a low to high carrier mobility ($\approx$2-510 cm$^2$ V$^{-1}$ s$^{-1}$)\cite{lee2014air,maughan2016defect,saparov2016thin,lopez2019optical,guo2017two}. The wide range of measured mobilities of Cs$_2$SnI$_6$ can be attributed to variations resulting from different synthesis and characterization methodologies. Additionally, significant discrepancies have been observed between theoretical and experimental results regarding the transport properties of this material.\cite{maughan2018anharmonicity,maughan2016defect,cai2017computational,bhumla2022vacancy} The intrinsic limitations to mobility in Cs$_2$SnI$_6$ are still not fully understood, and the underlying scattering mechanisms governing carrier transport remain elusive. Therefore, a comprehensive and systematic study encompassing both theoretical and experimental investigations is highly desired to unravel the mobility ambiguity in Cs$_2$SnI$_6$ and shed light on its transport characteristics.
As of now, this compound exhibits a PCE of only 2.1\%.\cite{lee2017solution} In contrast, substitutional alloying in Cs$_2$SnCl$_6$ yields high photoluminescence quantum yield (PLQY) of 95.4\% making it promising for further exploration in LED applications.\cite{tan2020lead} Despite considerable investigation into its structural, electronic, and optical properties, the elucidation of charge-carrier dynamics in Cs$_2$SnI$_6$ still poses challenges that hinder the optimization of conversion efficiencies.

     Similarly, Cs$_2$TiBr$_6$, Cs$_2$TeI$_6$, and  Cs$_2$PtI$_6$ are also studied experimentally for PV absorbers with their band gaps in the ideal visible range: 1.8, 1.5, and 1.4 eV respectively, along with high absorption coefficients ($\sim$10$^5$ cm$^{-1}$).\cite{vazquez2020vacancy,schwartz2020air} Here, device efficiency for Cs$_2$TiBr$_6$ as PV absorber is reported to be 3.3\%.\cite{chen2018cesium} Indirect band gap and material instability are reported to be responsible for poor PCE in this case. In another report, PV device with Cs$_2$PtI$_6$ shows PCE of 13.88\%, which is a remarkable improvement on the reported efficiencies among all the materials belonging to this class till date.\cite{schwartz2020air} Contribution of larger carrier lifetimes along with direct band gap in ideal visible range and robust stability  help Cs$_2$PtI$_6$ to attain the high PCE. There are reports of synthesizing Pd\cite{sakai2017solution}, and Zr\cite{abfalterer2020colloidal} based nanomaterials experimentally but not much has been explored  in the direction of optoelectronics. These background clearly indicates that A$_2$BX$_6$ class is extremely interesting and fertile from the application perspective, yet a detailed systematic study on their optoelectronic, carrier transport and phonon properties connecting these observations  is lacking. Moreover, it is also noticed that substitutional alloying/doping of pure material is an important strategy to improve optoelectronic properties, which again necessitates an in-depth understanding of the pure materials themselves.  
     
     In this communication, we present a detailed and systematic study on the A$_2$BX$_6$ class of materials by using highly accurate ab-initio calculations. First, we have performed a thorough stability analysis which includes choice of different structural prototypes, thermodynamical stability via chemical potential phase diagram and convex hull analysis, and lattice dynamics simulation. Next, we have studied the electronic properties of the stable compounds using hybrid (HSE06) functional, which is known to predict reasonably accurate electronic structure information. Optical absroption and PV device parameters are calculated on the promising set of systems showing band gaps in the ideal visible region. Finally, carrier transport properties of these compounds are studied by considering the important scattering mechanisms. The importance of electron-phonon interactions, calculated within the temperature-dependent
Feynman polaron model, is also discussed in some detail. We believe that such an in-depth study not only provides a solid physical basis on these class of semiconductors but will also be immensely beneficial for researchers working on their device application in the field of PV, LED, PD, and PC.

	\section*{II. Structural Properties and Stability}
	Vacancy ordered double perovskite, A$_2$BX$_6$ is a class of compounds where alternate BX$_6$ octahedra  are removed from the  ABX$_3$ unit cell as shown in Figure \ref{fig:1}(a). In other words, 50\% B cations are missing compared to the closed-packed A$_2$BB$'$X$_6$ perovskite structure. Here, A possesses +1 oxidation state, B has +4 oxidation state  and  X is halide anion with -1 oxidation state. In general, for perovskites, different crystal structures are possible depending on the ionic radii of the constituent elements. These structures are roughly dictated by few important geometrical factors as defined below, 
 \begin{itemize}
\item Goldschmidt's tolerance factor: \\ 
\hspace{6in }{$t=\left(\text{r}_\text{A}+\text{r}_\text{X}\right)/\sqrt{2}\left(\text{r}_\text{B}+\text{r}_\text{X}\right)$}
\item Octahedral factor : $\mu = \text{r}_\text{B}/\text{r}_\text{x}$
\item Radius ratio : $\text{r}_\text{A}/\left(\text{D}_\text{XX}- \text{r}_\text{X}\right)$
 \end{itemize}

In the above expressions, $\text{r}_\text{A}$, $\text{r}_\text{B}$, $\text{r}_\text{X}$ and $\text{D}_\text{XX}$ are the empirical ionic radii of the constituent elements $\text{A}$, $\text{B}$, $\text{X}$ and the nearest neighbour X-X bond length, respectively in the A$_2$BX$_6$ structure.  All the calculated parameters are tabulated in Table S1 of the supplementary information (SI).\cite{supplement}  The calculated Goldschmidt's tolerance factor predicts formation of cubic structures, which is also consistent with our stability analysis (discussed later) and experimental observations for few of the compounds reported in the  literature.\cite{ju2018earth,zhou2018all,schwartz2020air,maughan2016defect,saparov2016thin,sakai2017solution}

			In this work, we have investigated the following A$_2$BX$_6$ compounds: A= Cs; B= Ge, Te, Se, Sn, Pd, Pt, Ti, Zr, Hf; X=I, Br. 
    For each compound, we have considered seven most common structural prototypes (as reported in International Crystal Structure Database (ICSD))\cite{bergerhoff1983inorganic,icsd} for A$_2$BX$_6$ class of compounds. Space group of these seven structures are Fm-3m (cubic), I4$\slash$m (tetragonal), I4$\slash$mmm (tetragonal),  P-3m1 (hexagonal), Pnma (orthorhombic), P4$\slash$mnc (monoclinic), and P12$_1$$\slash$c1 (monoclinic). These crystal structures are shown in Fig. S1 of SI.\cite{supplement}  Most of these structures are very similar in symmetry and differ in energy only within a few meV (3-4 meV). Post structural optimization, the lowest energy structure for most of the above set of compounds turns out to be cubic (Fm-3m). It has been observed experimentally that several Cs based iodide and bromide compounds indeed crystallize in the cubic space group.\cite{ju2018earth,zhou2018all,schwartz2020air,maughan2016defect,saparov2016thin,sakai2017solution} 
   
   To further assess the chemical stability, we have calculated the convex hull energies (E$_\text{hull}$) of these compounds with respect to possible secondary phases available in ICSD, open quantum materials database (OQMD)\cite{kirklin2015open,saal2013materials} and materials project (MP) database\cite{WOS:000332272300003}. As evident from Fig. \ref{fig:1}(b), most of the compounds lie on the convex hull i.e. E$_\text{{hull}}$ = 0,  except Cs$_2$GeI$_6$, Cs$_2$SeI$_6$, Cs$_2$PdI$_6$, and Cs$_2$GeBr$_6$, confirming the stability of the former. For the remaining four compounds E$_\text{hull}$ ranges between 5-60 meV/atom, indicating likelihood of chemical (meta/in)stability. 

    Next, in order to explore the most probable secondary phases during synthesis of Cs$_2$BX$_6$ materials, we have calculated the compositional phase diagrams (chemical potential) for materials on the convex hull. More details about the phase diagram calculations/analysis are given in the Sec. S1(A) of SI.\cite{supplement}  Figure \ref{fig:1}(c-g) shows the phase diagrams for those materials which can be potential for optoelectronic applications (based on their band gap, optical and transport properties, as discussed later).  The phase diagrams for remaining compounds are displayed in Figure S2 of SI.\cite{supplement} The green shaded portion shows the stability regions of these materials. The extent of the stability region  directly correlates with the ease/difficulty of experimental synthesis.The theoretically optimized lattice parameters and bond lenghts (B-X) of all the stable compounds in their cubic structure are displayed in Table S2 of SI.\cite{supplement}

    Further, we have checked the dynamical stability of these compounds by calculating phonon dispersions as shown in Figure S3 of SI.\cite{supplement} The absence of any visible imaginary phonon modes indicates dynamical stability of these compounds.  For Cs$_2$SnI$_6$ and Cs$_2$TiI$_6$, one can observe small negative phonon frequencies the magnitude of which decreases with increasing supercell size. This is because the later captures the effect of higher order inter-atomic force constants more accurately. This is also evident in previously reported phonon dispersion for Cs$_2$SnI$_6$.\cite{bhui2022intrinsically,jong2019anharmonic} Nevertheless, these compounds are already experimentally synthesized, and hence naturally stable.\cite{maughan2016defect,liga2021colloidal}
 
  Following stability analysis, we have further studied the electronic structures of 15 (14 stable and 1 metastable) compounds in the next section.


			   \begin{figure*}[t!]
									\centering
									\includegraphics[width=1.0 \linewidth]{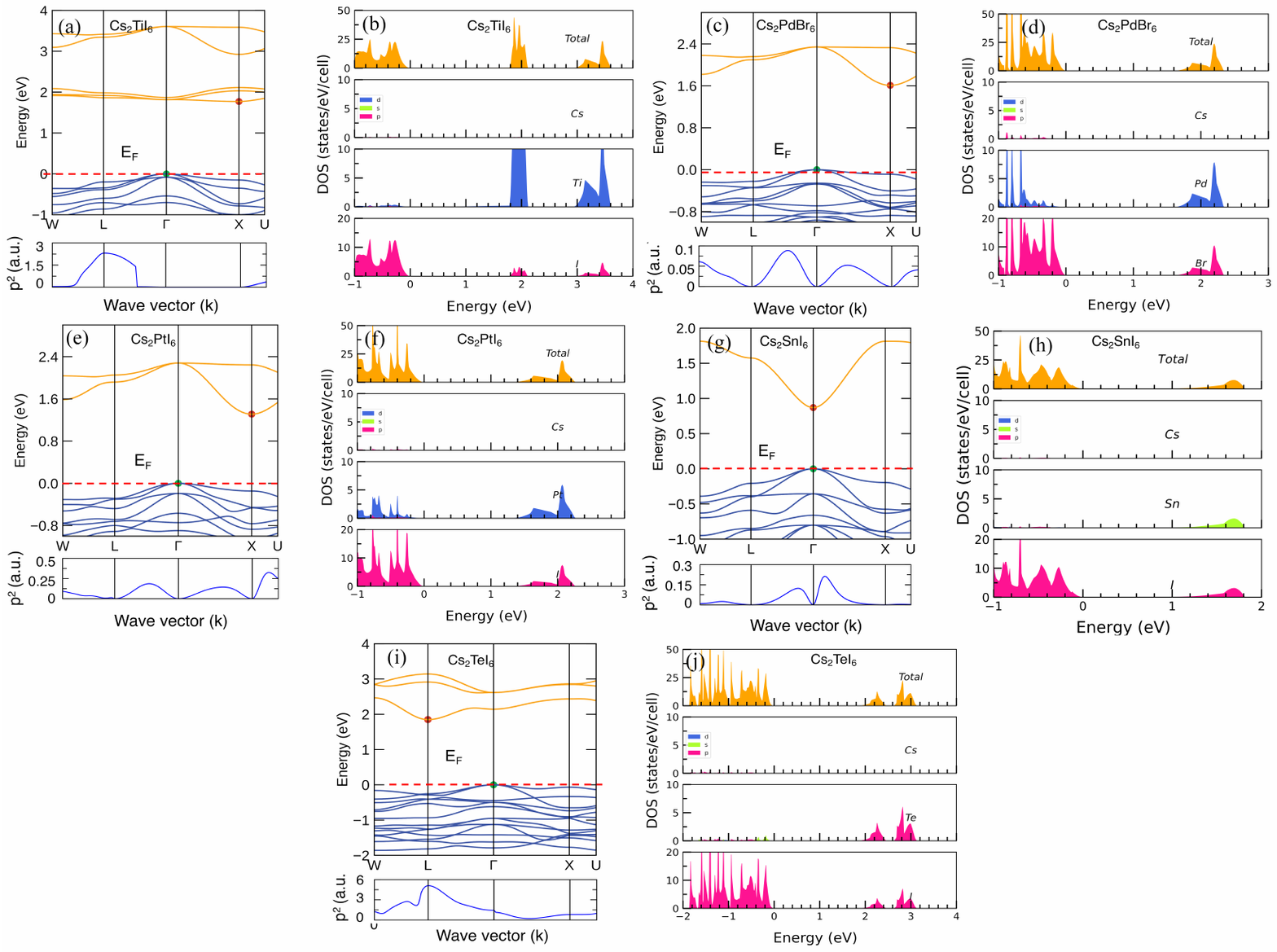} 
								\caption{Band structures and the square of dipole transition matrix elements (p$^2$) between VBM and CBM, for (a) Cs$_2$TiI$_6$, (c) Cs$_2$PdBr$_6$(e) Cs$_2$PtI$_6$, (g)Cs$_2$SnI$_6$, and (i) Cs$_2$TeI$_6$ respectively. (b), (d), (f), (h) and (j) show the projected density of states (PDOS) for the same set of compounds respectively. All the calculations are done using PBE functional including spin-orbit coupling (soc) effect while band gap is scissor shifted to HSE+soc calculated values. In band structure plots, VBM and CBM are indicated via green and red circles respectively.} 
									\label{fig:2}
								\end{figure*}

				\section*{III. Electronic structure }
				Band structure calculations for all the compounds are initially performed using  Perdew-Burke-Ernzerhof (PBE) exchange-correlation functional\cite{perdew1996generalized}.  As PBE functional is well-known to underestimate the band gap, we also employ hybrid Heyd-Scuseria-Ernzerhof (HSE06)\cite{krukau2006influence} functional which gives more accurate estimate of band gap  in comparison to experiment. Spin-orbit coupling (soc) effect in included in all the calculations. Band structures for four potential compounds calculated using PBE+soc functional (band gaps are scissor shifted to  HSE+soc values) are shown in Figure \ref{fig:2}. In A$_2$BX$_6$ class of compounds, the topology of band structure calculated using HSE+soc functional is very similar to that calculated using PBE+soc functional except the enlargement of band gap in the former (See Figure S4 of SI for few representative cases).

				Figure \ref{fig:2} also shows the optical transition probability, (square of dipole transition matrix elements, p$^2$), and total/orbital projected density of states (PDOS) of Cs$_2$TiI$_6$, Cs$_2$PdBr$_6$, Cs$_2$PtI$_6$, Cs$_2$SnI$_6$, and Cs$_2$TeI$_6$ respectively. The HSE06+soc band gap values for the respective compounds are provided in Table \ref{tab:t1}. The band structure, PDOS and respective band gap values for other compounds are provided in Figure S5-S7 and Table S3 and S4 of SI.\cite{supplement}  In Fm-3m phase, the estimated band gap values lie within 0.72 eV to 4.31 eV for different compounds. Optical transitions at the fundamental direct gaps are dipole forbidden for all the compounds, as confirmed by the calculated optical transition probability (p$^2$). Here the presence of inversion symmetry plays the key role to induce parity forbidden transitions for these compounds, effectively increasing the optical band gap.

           In the present study, we considered 9 different elements at the B site, belonging to 4 distinct groups in the periodic table. Despite all elements having a +4 oxidation state, their valence electron orbital configurations differ, resulting in distinct electronic structures, including variations in band structure and band gap types among the compounds. In the following, we shall discuss the electronic structure of representative compounds from each group and compare them with the electronic structures of other compounds within the same group, including different halides.
              
			     For Cs$_2$TiI$_6$, band gap is indirect in nature with conduction band minimum (CBM) at X and valence band maximum (VBM) at $\Gamma$ . But the direct band gap at $\Gamma$ is very close to indirect band gap value ($\sim$50 meV) (Table \ref{tab:t1}). From the orbital projected density of states (PDOS), we observe that CBM is comprised of Ti-d and I-p i.e. B-d and X-p (see Figure \ref{fig:2})(a,b)). The electronic band gap value is 1.77 eV which is overestimated by 0.75 eV with respect to experimental value (1.02 eV)\cite{ju2018earth}. The calculated optical band gap lies within  100 meV from the fundamental direct gap. Apart from that, the large difference between the calculated electronic band gap and optically measured experimental band gap can be attributed to the excitonic effect (not taken into account here) and the defects present in the measured sample, as discussed by B.Cucco et.al.\cite{cucco2021electronic}. All the electronic structure information for the rest of the compounds can be found in Figure S5-S7  and Table S3 and S4 of SI.\cite{supplement} It is clearly evident that the band gap increases from Ti $\rightarrow$ Zr $\rightarrow$ Hf and also with I $\rightarrow$ Br. In this group, only Cs$_2$TiI$_6$ shows band gap in the ideal visible region.
			     
			    Cs$_2$PdI$_6$ shows indirect band gap in both the space groups with CBM at X and VBM at $\Gamma$ (see \textcolor{black}{Fig. S5} of SI). The optically allowed direct band gap (0.88 eV) is very close to the indirect band gap values (0.72 eV) (shown in Table \ref{tab:t1} ). Experimentally, Cs$_2$PdI$_6$ nanocrystals\cite{zhou2018all} are synthesized and and a band gap of 0.69 eV is reported. The reason behind the overestimation might be similar to what is explained for the case of  Cs$_2$TiI$_6$. In this case, the CBM  is comprised of Pd-d, I-p orbitals  while VBM is composed of only I-p orbital (see \textcolor{black}{Fig. S5} of SI).  Like Cs$_2$PdI$_6$,  Cs$_2$PtI$_6$, Cs$_2$PtBr$_6$, and Cs$_2$PdBr$_6$ show similar orbitals contribution at both CBM and VBM giving rise to indirect nature of band gap. Their band gap values along with the formation energetics and different between direct and indirect band gaps are presented in SI (see \textcolor{black}{Tables S3 and S4} and \textcolor{black}{Fig. S5 and Fig. \ref{fig:2}(c,e)}. For Cs$_2$PtI$_6$ and Cs$_2$PdBr$_6$, the calculated band gap is close to experimentally reported values of  Cs$_2$PtI$_6$ powder \cite{hamdan2020cs2pti6, yang2020novel} and Cs$_2$PdBr$_6$ nanocrystals \cite{zhou2018all,sakai2017solution} respectively. Here, we observe an increase in band gap going from Pd $\rightarrow$ Pt and also from I $\rightarrow$ Br. In this case, Cs$_2$PdI$_6$, Cs$_2$PtI$_6$, and Cs$_2$PdBr$_6$ compounds show band gaps within the ideal visible range.
			     
			     The band structure analysis of Cs$_2$TeI$_6$ reveals that it has indirect band gap with a value of 1.85 eV, consistent with the study by Maughan et.al.\cite{maughan2016defect}. From PDOS analysis, we observe that the CBM is comprised of Te-p and I-p orbitals whereas VBM is made up of I-p orbital (see Figure \ref{fig:2}(i) and (j)).  The calculated electronic band gap value of 1.85 eV is 0.26 eV higher than the experimentally reported value \cite{maughan2016defect}. For Cs$_2$TeBr$_6$, the band gap nature and orbital contribution at both CBM and VBM is similar to that of  Cs$_2$TeI$_6$. The related electronic properties can be found in SI (see \textcolor{black}{Table. S4} and \textcolor{black}{Fig. S6} ). All the electronic structure information for Cs$_2$SeBr$_6$ can be found in SI (see Table S4 and Figure S6 (c,d) ), which shows similar orbital characteristics.

			     For Cs$_2$SnI$_6$, the calculated band gap value is 0.85 eV which is direct in nature with band edges (at $\Gamma$) in agreement with the values reported by Maughan et.al.\cite{maughan2016defect} This is 0.38 eV higher than the experimentally reported value \cite{maughan2016defect}. From orbital analysis, we observe that the CBM is made up of Sn-s and I-p orbitals, and VBM is comprised of I-p orbital (see \textcolor{black}{Fig. 2(g,h)}). For Cs$_2$SnBr$_6$, the band gap nature and orbital contribution at both CBM and VBM remains similar to that of Cs$_2$SnI$_6$. The related electronic properties can be found in SI (see \textcolor{black}{Table S4} and \textcolor{black}{Fig. S6(a,b)}). 

\begin{figure*}[t!]
						     			\centering
						     			\includegraphics[width=1.0 \linewidth]{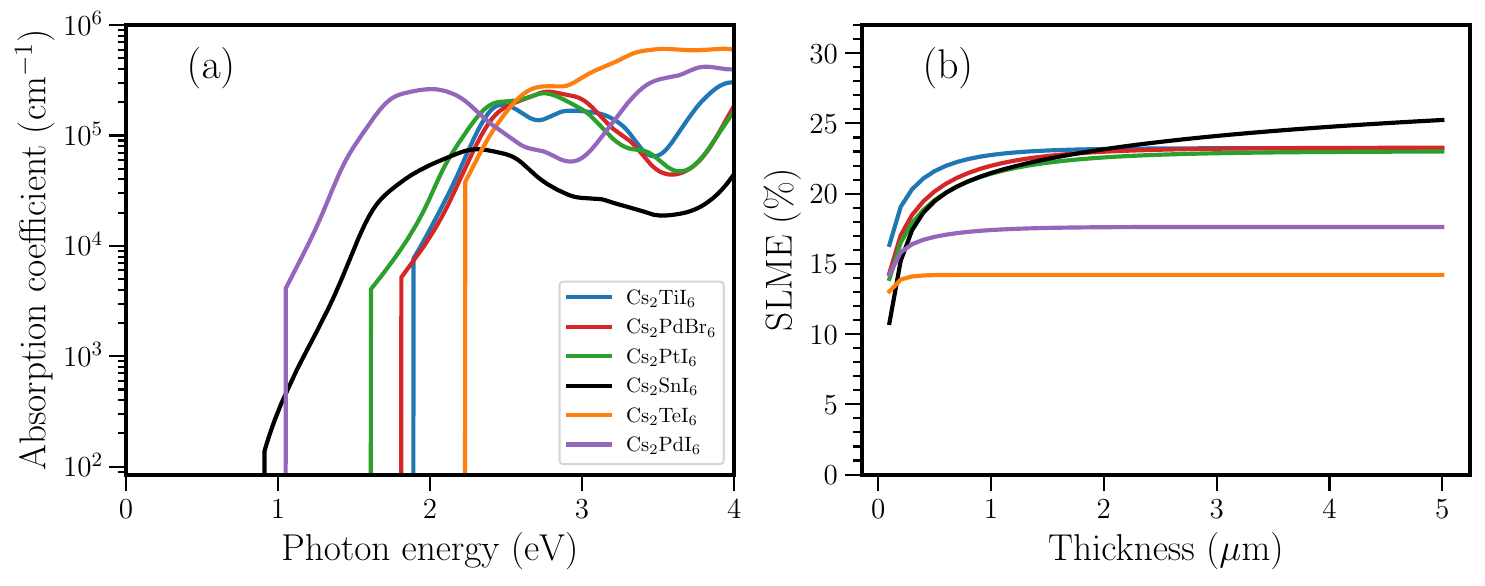}
						     			\caption{ (a) Absorption coefficients vs incident photon energy and (b) spectroscopic limited maximum efficiency (SLME) (298 K) vs. film thickness  for Cs$_2$TiI$_6$, Cs$_2$PdBr$_6$, Cs$_2$PtI$_6$, Cs$_2$SnI$_6$, Cs$_2$TeI$_6$, and Cs$_2$PdI$_6$ compounds. }
						     			\label{fig:3}
						     		\end{figure*}
			     
			     To summarize the electronic properties, one should note that the calculated electronic band gaps are always overestimated as compared to the experimentally reported optical band gap which is a well-known fact.\cite{cucco2021electronic} Contrary to previous reports,  the optical band gaps are close to the lowest direct band gaps, confirmed by our calculation of optical transition probability. We believe that  the most probable reasons for the theoretical overestimation of band gaps can be attributed to the excitonic effects (not included in the present calculations) and defects present in the experimental samples, as discussed by B.Cucco et.al.\cite{cucco2021electronic}.

			     From the electronic structure analysis, we notice that the band gaps of  Cs$_2$TeI$_6$, Cs$_2$SnI$_6$, Cs$_2$PdI$_6$ (Pnma), Cs$_2$PtI$_6$, Cs$_2$TiI$_6$, and Cs$_2$PdBr$_6$ lie in the ideal visible range for photovoltaic application. Therefore, we shall now focus on the optical properties of these six compounds along with the well-known descriptor ‘Spectroscopic limited maximum efficiency’ (SLME) (as proposed by Yu et.al\cite{yu2012identification}) to better understand their potential as solar absorber.

  \begin{table*}[t!]
  \small
  		\begin{centering} 
  			\begin{tabular}{c c c c c c c c c c c}
  				
  				\hline
  				\hline
  				Compound & $E_g^{(expt)}$ &$E_g$(HSE+soc)& $\Delta$$E_g^{da}$&J$_{SC}$ & J$_{max}$ & V$_{OC}$ & V$_{max}$ & FF&  SLME  \tabularnewline
  				\vspace{0.02 in}  
  				& (eV)  & (eV) & (meV) & (mA cm$^{-2}$)   & (mA cm$^{-2}$) & (V) & (V) & & ($\eta \%$) \tabularnewline
  				
  				\hline 
  				\hline \\
  				Cs$_2$TiI$_6$  &  1.02\cite{ju2018earth}& 1.77 (ID)& 72 & 16.64 & 16.34 &1.50 & 1.39 & 0.91 &22.78  \\
  				\\ \hline \\
                  Cs$_2$PdBr$_6$  & 1.6\cite{sakai2017solution}, 1.69 \cite{zhou2018all}&  1.61 (ID) & 110 &17.62 & 17.26& 1.35 & 1.25  & 0.91 &21.63 \\								
  				\\ \hline \\
  				
  				Cs$_2$PtI$_6$ &1.25\cite{hamdan2020cs2pti6}, 1.37 \cite{yang2020novel}, 1.4 \cite{schwartz2020air} &   1.31 (ID)  &  149 & 22.23 & 21.66 & 1.08 & 0.98  & 0.89& 21.26 \\
  				\\ \hline \\
  												
  				Cs$_2$SnI$_6$ &1.25 \cite{maughan2016defect}, 1.62 \cite{saparov2016thin} &0.87 (D)& 41 & 32.58& 31.33 & 0.72 & 0.64   & 0.85 & 20.07 \\
  				\\ \hline \\
  				Cs$_2$TeI$_6$  & 1.59 \cite{maughan2016defect}& 1.85 (ID)& 190 & 12.95  & 12.72 & 1.55 & 1.45  & 0.92 & 18.44 \\
  				\\ \hline \\
  				Cs$_2$PdI$_6$ & 1.41 \cite{zhou2018all}  &0.72 (ID) & 166 & 43.20 &  40.94 & 0.54 & 0.46  & 0.81 & 18.97 \\
  				\\ \hline 
  				\hline  
  			\end{tabular}
  			\par\end{centering}
  	\caption{Simulated band gap ($E_g$), difference between electronic and optically allowed direct band gap ($\Delta$$E_g^{da}$), short-circuit current density (J$_{SC}$ ), open-circuit voltage (V$_{OC}$ ), current density (J$_{max}$ ) and voltage (V$_{max}$ ) at maximum power, spectroscopic limited maximum efficiency (SLME), and fill factor (FF) for Cs$_2$TiI$_6$, Cs$_2$PdBr$_6$ Cs$_2$PtI$_6$ Cs$_2$SnI$_6$, Cs$_2$TeI$_6$, and Cs$_2$PdI$_6$ compounds. ID and D indicates indirect and direct nature of band gaps respectively. All the device-related parameters are shown for 500 nm thickness at 298K. Experimental band gaps ($E_g^{(expt)}$) are also listed for comparison.  }
  	\label{tab:t1}
  \end{table*}
 
	
		\section*{IV. Optical Properties}
		
		Figure \ref{fig:3}(a) shows the absorption coefficients for the above mentioned six promising compounds. All these compounds  can act as potential solar absorber as their absorption coefficients are as high as 10$^{4}$-10$^{5}$ cm$^{-1}$ in the visible range. The optical absorption is contributed by two factors: (1) optical joint density of states (JDOS) and (2) optical transition strength. As we can see in Figure S5(a-d),\cite{supplement} the square of dipole transition matrix elements (aka transition strength) for Cs$_2$PdI$_6$ is pretty high contributing to better optical absorption. This can be attributed to Cs-p, I-p to Pd-d transition. Apart from that, the JDOS is also likely to be high as the bands near CBM and VBM show flat nature. They are comprised of `p' and `d' orbitals, showing more localized nature as compared to the other compounds. In addition to Cs$_2$PdI$_6$, Cs$_2$SnI$_6$, and Cs$_2$TeI$_6$ also show similar absorption coefficient spectrum. The absorption coefficient is directly related to the frequency dependent dielectric function  of a semiconductor via following equation
		
		\begin{eqnarray}
		\alpha(E)=\frac{2\omega}{c}\sqrt{\frac{\sqrt{\epsilon^2_{re}+\epsilon^2_{im}}-\epsilon_{re}}{2}}
		\end{eqnarray}
		
		where $E$ is the incident photon energy, $\omega$ is the angular frequency related to $E$ via $E=\hbar\omega$, $c$ is the velocity of light, $\epsilon_{re}$ and $\epsilon_{im}$ are the real and imaginary part of the dielectric function respectively. Figure \ref{fig:3}(b) shows the thin-film thickness dependance of spectroscopic limited maximum efficiency (SLME), which turn out to be >15\% for all six compounds. Interestingly, we can see a higher SLME for the  Cs$_2$PdI$_6$, Cs$_2$PtI$_6$ and Cs$_2$TiI$_6$ compounds as compared to $Cs_2SnI_6$. Such increased SLME is essentially attributed to increased high absorption spectra due to I-p to Pd/Pt/Ti-d orbital transition as well as suitable band gaps. In Table \ref{tab:t1}, we present the simulated device parameter values for the six compounds: short-circuit photo-current density (J$_{\mathrm{sc}}$), open circuit voltage (V$_{\mathrm{oc}}$), fill factor ($FF$), maximum current density (J$_{\mathrm{max}}$), and maximum voltage (V$_{\mathrm{max}}$) obtained from the SLME calculation. The detailed description and method of calculation of these parameters can be found in the SI.\cite{supplement} As expected, materials with higher band gaps exhibit higher V$_{oc}$ values, while materials with lower band gaps acquire higher J$_{sc}$ values. The other compounds do not have SLME as high as these six materials owing to higher band gaps. Their absorption coefficients and SLME values are shown in Figure S8 and S9 of SI respectively.\cite{supplement} Furthermore, it is worth noting that the band gaps of other compounds are distributed within the visible range (larger than 1.8 eV), which makes them suitable for LED and photocatalytic water splitting applications.  Alloying at B and x sites are other avenues to tune the optoelectronic properties of these systems and hence make them suitable for different applications.

	\section*{V.  Transport properties}

                                    \begin{figure*}
						     			\centering
						     			\includegraphics[width=1.0 \linewidth]{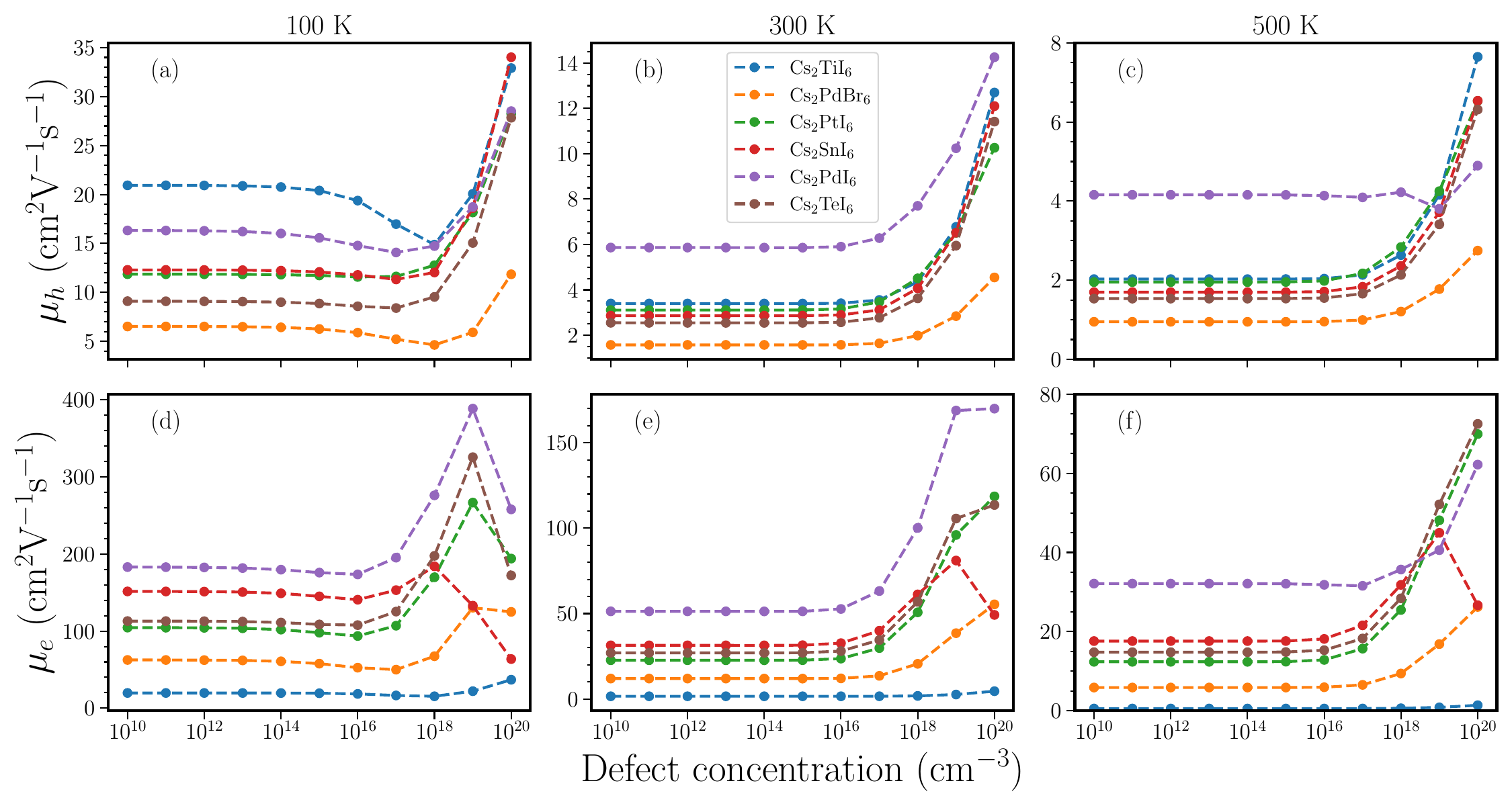}
							     			\caption{ (a,b,c) Hole mobility ($\mu_h$) and (d,e,f) electron mobility ($\mu_e$)  for Cs$_2$TiI$_6$, Cs$_2$PdBr$_6$, Cs$_2$PtI$_6$, Cs$_2$SnI$_6$, Cs$_2$PdI$_6$, and Cs$_2$TeI$_6$ compounds  as a function of defect concentrations at three different temperatures, T= 100K, T=300 K and T= 500K respectively.  }
						     			\label{fig:4}
						     		\end{figure*}

						     		\begin{figure*}[t!]
						     			\centering
						     			\includegraphics[width=1.0 \linewidth]{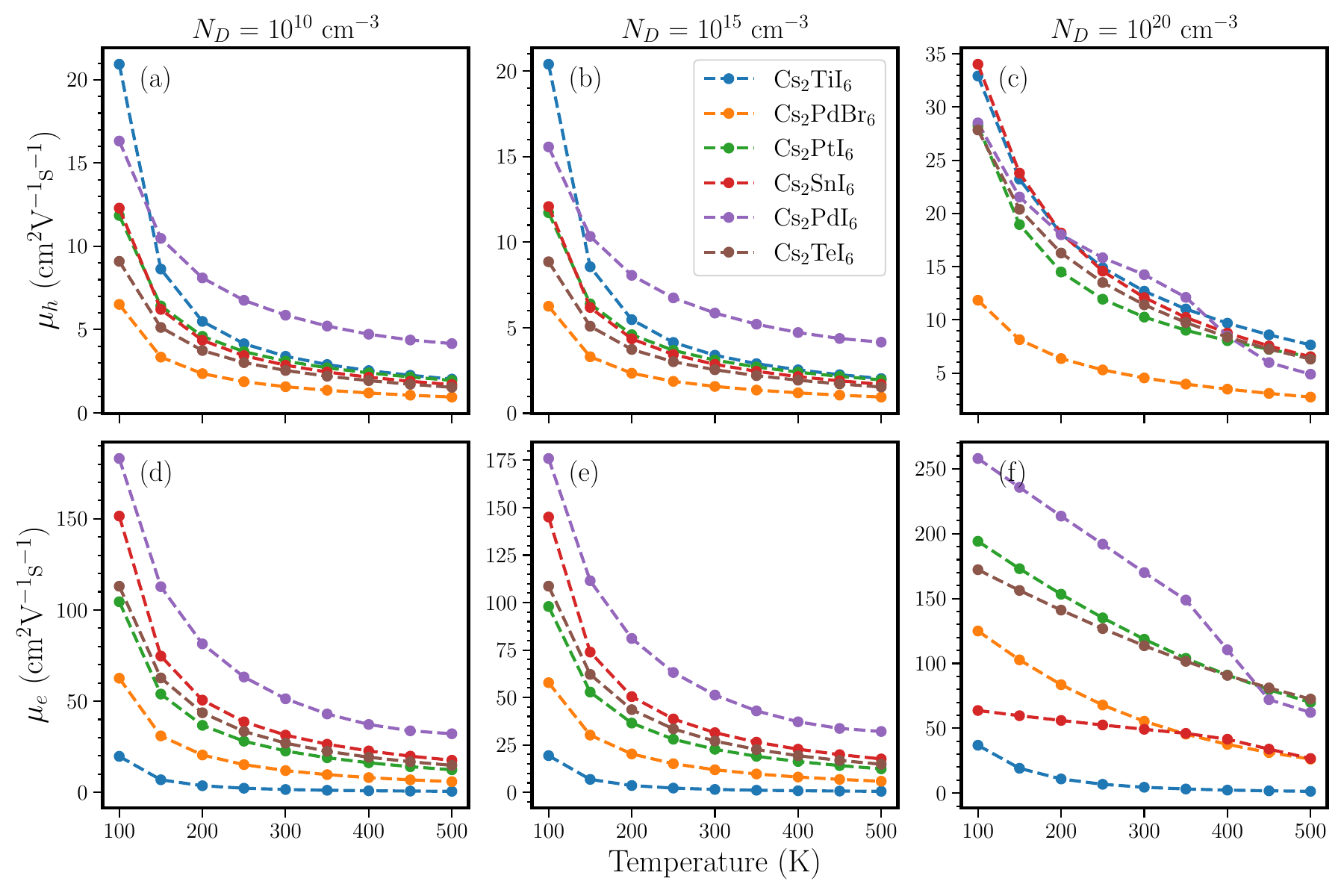}
						     			\caption{ (a,b,c) Hole mobility ($\mu_h$) and (d,e,f) electron mobility ($\mu_e$)  for Cs$_2$TiI$_6$, Cs$_2$PdBr$_6$, Cs$_2$PtI$_6$, Cs$_2$SnI$_6$, Cs$_2$PdI$_6$, and Cs$_2$TeI$_6$ compounds as a function of temperatures at three different defect concentrations, N$_D$=10$^{10}$ cm$^{-3}$, 10$^{15}$ cm$^{-3}$, and 10$^{20}$ cm$^{-3}$  respectively.}
						     			\label{fig:5}
						     		\end{figure*}

             Detailed analysis of optoelectronic properties and calculation of solar efficiency (SLME)  reveals six compounds to be promising. This can be attributed to their optimal band gaps falling within the ideal visible region of the solar spectrum coupled with their excellent absorption coefficients.  However,  in a practical photovoltaic device, extraction of charge carriers is one of the key component determining its power conversion efficiency. As such, mobility of the charge carriers is an integral quantity dictating the promise of a semiconductor for solar harvesting. Most of the past theoretical studies on photovoltaic materials rely on  calculation of transport properties based on constant relaxation time approximation (RTA). Within this approximation, all the scattering mechanisms are averaged out via a single relaxation time (chosen to be approximately 10 fs). This practice, however, can be misleading as the carrier relaxation time is a complex parameter which sensitively depends on a number of physical properties and can be significantly different for different materials belonging to the same class (as illustrated in this study). In this section, we perform a thorough analysis of the carrier mobilities of these compounds considering three relevant scattering mechanisms, namely, acoustic phonons (ADP), ionized impurities (IMP), and polar optical phonons (POP) scattering. We have excluded piezoelectric scattering due to the inherent centro-symmetry present in these compounds. In Figure \ref{fig:4} and \ref{fig:5}, we show the temperature and defect concentration dependence of electron and hole mobilities ($\mu_e$ and $\mu_h$) for these compounds. Contribution of individual scattering mechansims on these mobilities for the six compounds are provided in Figure S10 to S15 of SI.\cite{supplement} 


Figure S16 of SI\cite{supplement} displays the total relaxation times of six compounds at varying defect concentrations, ranging from 10$^{10}$ cm$^{-3}$ to 10$^{20}$ cm$^{-3}$ at three different representative temperatures (100 K, 300 K, and 500 K) for both hole and electron transport. For defect concentrations in the low to moderate range, the relaxation times remains almost constant. However, as the defect concentration increases, relaxation times vary in an irregular manner. To comprehend the cause of this anomalous behavior, a more in-depth analysis was conducted. The relaxation times for all three scattering mechanisms were calculated for each compound, and plotted in Figures S17 to S22.\cite{supplement} A close inspection of these data confirms that in the low to moderate concentration range, the primary scattering mechanism is due to POP scattering. In contrast, as the concentration increases into the higher range, the dominant scattering mechanism shifts to IMP scattering, resulting in the emergence of anomalous behavior. Such unusual behavior is also reflected in the mobility shown in Fig. \ref{fig:4}.

Speaking about the behavior of mobilities for each individual compounds, one can notice that at low temperature (100 K), the hole mobility ($\mu_h$) is highest for Cs$_2$TiI$_6$ ($\sim$20.9 cm$^2$V$^{-1}$ s$^{-1}$). With increasing temperature, $\mu_h$ decreases slowly to reach a shallow minimum and increases again with increasing defect concentration. At higher temperature, Cs$_2$PdI$_6$ shows the highest hole mobility ($\sim$5.9 cm$^2$V$^{-1}$ s$^{-1}$ @300 K and $\sim$4.2 cm$^2$V$^{-1}$ s$^{-1}$ @500 K ) among all the compounds.  This compound also shows the highest electron mobility ($\sim$183 cm$^2$V$^{-1}$ s$^{-1}$ @ 100 K, $\sim$51 cm$^2$V$^{-1}$ s$^{-1}$ @ 300 K, and $\sim$32 cm$^2$V$^{-1}$ s$^{-1}$ @ 500 K) throughout the temperature range.  At room temperature, the hole mobilities remain relatively low but except Cs$_2$TiI$_6$, electron mobilities show moderate to high values ($\sim$13-63 cm$^2$V$^{-1}$ s$^{-1}$). This is commensurate with the electronic band structures of these compounds where the VBM shows flat bands  whereas the CBM is more dispersive.  Consequently, n-type doping could prove advantageous for efficient charge carrier collection in photovoltaic devices, aligning with the experimental findings for this class of compounds.\cite{maughan2018tolerance,lee2014air,saparov2016thin}  A closer look at the individual contributions from the different scattering mechanisms show that at low to moderate defect concentrations ($< 10^{18}$ cm$^{-3}$), POP scattering is the dominant scattering mechanism limiting the mobilities. With increasing temperatures, the number of activated polar optical phonons increase, and as a result we see a decrease in overall mobility going from 100 K $\rightarrow$ 300 K $\rightarrow$ 500 K. At higher defect concentrations( 10$^{18}$-10$^{20}$ cm$^{-3}$), we see ionized impurity scattering begins to dominate as can be seen from Figures S10(a,b,c)-S15(a,b,c) of SI.\cite{supplement} At these concentrations, there is one more mechanism that starts to impact the carrier mobility, which is the screening of polar optical phonons by free carriers. This in effect reduces the POP scattering, effectively increasing the overall mobility in some cases. Now, the temperature has also an effect on this screening mechanism. At higher temperatures, there are more activate polar optical phonons, which require a higher density of free carriers to effectively screen the Coulomb field created by these phonons. This is clearly evident from our SI plots (see Figures S10(a,b,c)-S15(a,b,c)).\cite{supplement} In all the cases, ADP scattering remains low which is common in  hybrid perovskites arising out of small deformation potentials.\cite{ganose2021efficient,irvine2021quantifying}

             In Figure \ref{fig:5}(a-f), we show average hole and electron mobilities with respect to temperatures ranging from 100 K to 500 K for three different defect concentrations, low (10$^{10}$ cm$^{-3}$), moderate (10$^{15}$ cm$^{-3}$) and high (10$^{20}$ cm$^{-3}$). Due to weak dependence on IMP scattering in low to moderate defect concentrations, we see the carrier mobility remains similar in these two defect concentrations. But we can see that at higher concentrations, IMP starts to dominate. As such, controlling the defect concentrations can impact device efficiencies, not only because at higher defect concentrations, IMP becomes the dominant scattering mechanism, but also because the prevalence of free carriers will start to screen the POP scattering effect.  As expected, the overall mobility has a strong temperature dependence for most of the compounds and remains high to moderate for the electrons whereas the hole mobility values remain consistently low.

The above analysis reveals that in A$_2$BX$_6$ class,  polar optical phonons play a dominant role at the realistic defect concentrations relevant for photovoltaic application. As such, next we study the properties of the polaronic states via calculating the Fr$\ddot o$hlich interactions under the temperature-dependent Feynman polaron model.\cite{feynman1955slow}\textcolor{black}{[Reference]} In polar semiconductors, for example, halide perovskite and its derivatives, the interaction between charge carriers and the macroscopic electric field generated by longitudinal optical phonon (LO) is well known to be the dominant scattering mechanism near room temperature which is expected to be the case for our studied materials as well.\cite{hellwarth1999mobility,manna2020lattice,maughan2018anharmonicity,maughan2018tolerance,ganose2021efficient} To investigate the same, we studied the influence of changing B-site in A$_2$BX$_6$ 
 on the electron-phonon coupling (EPC). Within the Fr$\ddot o$hlich interaction model, the interaction strength ($\alpha$) is defined as     
	    \begin{eqnarray}
	     	\alpha=\frac{1}{4\pi\epsilon_0}\frac{1}{2}\left(\frac{1}{\epsilon_{\infty}}-\frac{1}{\epsilon_{static}}\right)	\frac{e^2}{\hbar\omega_{LO}} \left(\frac{2m^*\omega_{LO}}{\hbar}\right)^{1/2}	
	    \end{eqnarray}
	     	    were, $\epsilon_0$ is dielectric constant of vacuum, $\epsilon_{\infty}$ and $\epsilon_{static}$ are high frequency and static dielectric constants of the  semiconductor, $\hbar$ is the reduced Plank constant, $\omega_{LO}$ is the characteristic angular LO frequency where all the infrared active optical phonon branches are taken into account via a spectral average,\cite{hellwarth1999mobility} $m^*$ is the carrier effective mass. Table \ref{tab:t2} display all the associated values related to Fr$\ddot o$hlich interaction for electrons for the six compounds. The corresponding list of parameters for the holes for these six compounds are reported in Table S5 of SM.\cite{supplement}
           To validate our simulation, we compare the calculated values of $\alpha$ for Cs$_2$SnI$_6$, with recent literature and observe a fair agreement.\cite{maughan2018anharmonicity}  In case of A$_2$BX$_6$ class, calculated $\alpha$-values lie in the moderate range (1$ < \alpha <$ 6). Estimated values of polaron radius (l$_p$) indicates formation of large polarons, similar to what is observed for hybrid halide perovskites and double perovskites.\cite{hellwarth1999mobility,manna2020lattice,saxena2020contrasting}  $\alpha_e$ value is highest for Cs$_2$TiI$_6$ mainly due to higher electron effective mass compared to the other compounds. Additionally, taking an inference from the electronic structure of these materials, we see that CBM in Cs$_2$TiI$_6$  has a contribution from Ti-d and I-p orbitals whereas for Cs$_2$SnI$_6$, it is Sn-s and I-p orbitals. Now, Ti-d orbitals are more localized arising out of the flat band and hence higher effective mass.  For other compounds, we see more dispersive bands (see Figure  \ref{fig:2}) at CBM, and the corresponding $\alpha$ values are in the range close to that of Cs$_2$SnI$_6$. Interestingly, the hole mobility turn out to be significantly lower than the electron mobility.  To conclude, large polaron is the main carrier related to moderate mobility for our studied compounds. \textcolor{black}{These crucial observations clearly indicates the importance of studying charge carrier behavior in A$_2$BX$_6$ class of compounds and its implications in future applications.}

	   		\begin{table}[t]
	   				\resizebox{\columnwidth}{!}{
	   			\small
	   			\begin{centering} 
	   				\begin{tabular}{c c c c c c c   }
	   					
	   					\hline
	   					\hline
	   					\tabularnewline
	   					\vspace{0.02 in}  
	   					Parameters & Cs$_2$TiI$_6$ & Cs$_2$PdBr$_6$ & Cs$_2$PtI$_6$ &  Cs$_2$SnI$_6$ & Cs$_2$TeI$_6$ & Cs$_2$PdI$_6$\tabularnewline [0.1 in]								
	   					\hline 
	   					\hline  \\ 	   					
	   					m$_e$$^*$(m$_0$)  & 1.66   &0.66  &0.49 & 0.31  & 0.35 & 0.61	\\  [0.03 in]
	   					
	   					
	   					$\epsilon_\infty$  &	4.91   & 4.15 & 4.82 & 5.04  & 4.89 &  7.58 \\ [0.03 in] 
	   					$\epsilon_{static}$& 10.81  & 7.21 & 8.40 & 10.41  & 11.21 & 11.68\\  [0.03 in]
	   					$\nu_{eff}$(THz)   &3.77    & 3.16  & 2.16 & 2.60  & 2.71 & 2.25	 \\  [0.03 in]
	   		
	   					$\alpha_e$         & 4.23   & 2.66 &	2.44 & 2.02   & 2.38 & 1.38 \\  		[0.03 in]	
                        $\mu_e$ (cm$^2$V$^{-1}$ s$^{-1}$) & 1.5 & 11.9  & 22.7 & 31.4   & 27& 51.3 \\  		[0.03 in]	
	   					
	   					l$_p(e)$(\AA)        &	29.40  & 32.75 & 25.88 &   32.48  & 30.92& 35.33 \\ [0.01 in]
	   					\\			  
	   					\hline 
	   					\hline  
	   				\end{tabular}
	   				\par\end{centering}}
	   			\caption{ Calculated electron effective mass (m$_e$$^*$ in units of electron rest mass (m$_0$)), high frequency and static dielectric constants ($\epsilon_\infty$ and $\epsilon_{static}$), effective  phonon frequency ($\nu_{eff}$), Fr$\ddot o$hlich coupling constants for electrons ($\alpha_e$), electron mobility ($\mu_e$), 
 and polaron radius (l$_p$)  for Cs$_2$TiI$_6$,  Cs$_2$PdBr$_6$,  Cs$_2$PtI$_6$ Cs$_2$SnI$_6$, Cs$_2$TeI$_6$ and Cs$_2$PdI$_6$ compounds. The corresponding set of parameters for holes are displayed in Table S5 of SI.\cite{supplement}}
	   			\label{tab:t2}
	   		\end{table}

				\section*{VI. Conclusion}
				In summary, we performed an accurate and systematic investigation of Pb-free vacancy ordered double perovskites (A$_2$BX$_6$) from the optoelectronic application perspective. We carried out a thorough stability analysis considering different structural prototypes and carefully simulating the convex hull energy diagram including all  possible secondary phases. We found 14 compounds to be stable and 1 in the metastable phase. For stable compounds, we further simulated the compositional phase diagrams to assist the experimentalists identifying the most probable secondary phases which might emerge during synthesis. Next, a carefull electronic structure analysis reveals six compounds, namely Cs$_2$TeI$_6$, Cs$_2$SnI$_6$, Cs$_2$PdI$_6$, Cs$_2$PtI$_6$, Cs$_2$TiI$_6$, and Cs$_2$PdBr$_6$ to possess optically allowed band gaps in the ideal visible range \textcolor{black}{(0.8-1.85 eV)}. The detailed investigation of optical properties confirms that few of these compounds possess favorable optoelectronic properties facilitating better efficiency than some of the existing ones. A close inspection of transport properties reveals that Cs$_2$PdBr$_6$,  Cs$_2$PtI$_6$, Cs$_2$SnI$_6$, Cs$_2$PdI$_6$, and Cs$_2$TeI$_6$ compounds acquire moderate to high electron mobilities ($\sim$13 - 63 cm$^2$V$^{-1}$ s$^{-1}$). In all the cases, polar optical phonons (POP) remain the dominant scattering mechanism at low to moderate defect concentrations. At high defect concentrations, ionized impurity scattering starts to dominate while accumulation of free carriers shows a screening effect on the POP scattering. This study is expected to facilitate the necessary base and guidance  for future experimental synthesis of some of these compounds to achieve desired features for promising device applications.

			\section*{VII. Computational Details} 	
		  First-principles calculations are carried out using density functional theory (DFT)\cite{PhysRev.140.A1133} with projector augmented wave (PAW)\cite{blochl1994projector}  basis set as implemented in  Vienna Ab-Initio Simulation Package (VASP).\cite{PhysRevB.59.1758,kresse1993ab,kresse1994ab,kresse1996efficient,KRESSE199615} A plane wave energy cutoff of 520 eV,  $\Gamma$-centered 4$\times$ 4$\times$4 k-mesh, and Perdew-Burke-Ernzerhof (PBE) exchange-correlation functional\cite{perdew1996generalized} were employed to perform the geometry optimization. The crystal structure was relaxed with force tolerance criteria of 0.001 eV\r{A}$^{-1}$. The spin{-}orbit coupling (soc) effect is included while simulating the electronic and optical properties. Hybrid(HSE06) functional\cite{krukau2006influence} is used to calculate the band gap and band edges which are known to provide a more accurate estimate for the same. Optical absorption spectra are simulated within the independent particle approximation and then the absorption onset value is scissor shifted to HSE06 band gap values. This method makes it possible to accurately assess the SLME for the materials under consideration. The chemical phase diagrams are drawn using Chesta software package.\cite{chesta}  Phonon dispersion is calculated using the density functional perturbation theory (DFPT) using $\Gamma$-centered 4$\times$4$\times$4 k-mesh under the supercell method. The 2nd order force constant is calculated using 2$\times$2$\times$2 supercells of primitive cells for cubic structures and in similar proportion for other structures.  Next, the rotational sum rule is applied using the hiphive package\cite{eriksson2019hiphive} to renormalize the phonon frequencies. Transport calculations are performed using the AMSET code,\cite{ganose2021efficient}, where we have considered three different scattering mechanisms, namely scattering due to acoustic phonons (ADP), ionized impurities (IMP), and polar optical phonons (POP). Piezoelectric scattering is not included due to the centro-symmetric crystal structure of A$_2$BX$_6$, whereas screening due to free carriers at high defect concentrations is included. This program uses the Boltzmann transport equation's (BTE) momentum relaxation time approximation (MRTA) to determine scattering rates and carrier mobilities. Polaron related parameters were simulated via implementing a temperature dependent Feynman polaron model.\cite{hellwarth1999mobility, frost2017calculating}  Born effective charges and static and high-frequency dielectric tensors were calculated using density functional perturbation theory (DFPT) as implemented in VASP. The effective mass has been calculated using the following equation:
        \begin{equation}
		  m^{*}=3\left[\frac{1}{m_{xx}^{*}}+\frac{1}{m_{yy}^{*}}+\frac{1}{m_{zz}^{*}}\right]
		  \end{equation} 

    where, $m_{ii}^{*}$ is the effective mass in the $i$-th direction ($i$=x,y,z).\cite{mortensen2005real,enkovaara2010electronic,larsen2017atomic,kangsabanik2022indirect}

		\section*{VIII. Acknowledgments}
	 SG acknowledges financial support from IIT Bombay for research fellowship. AA and MA  acknowledges National Center for Photovoltaic Research and Education (NCPRE) funded by Ministry of new renewable energy (MNRE), Government of India, and IIT Bombay for possible funding to support this research.

	\bibliographystyle{unsrt}
	\bibliography{ref.bib}
\end{document}